\documentclass[letterpaper,12pt]{article}
\usepackage{tabularx} 
\usepackage{amsmath}  
\usepackage{graphicx} 
\usepackage[margin=1in,letterpaper]{geometry} 
\usepackage{cite} 
\usepackage[final]{hyperref} 
\usepackage{amssymb}
\usepackage{hyperref}
\usepackage{tikz-feynman}
\usepackage{authblk}
\tikzfeynmanset{compat=1.1.0}
\usepackage{booktabs}
\usepackage{multirow}
\usepackage{subcaption}
\usepackage{setspace}   
\usepackage{indentfirst}
\hypersetup{
	colorlinks=true,       
	linkcolor=blue,        
	citecolor=blue,        
	filecolor=magenta,     
	urlcolor=blue         
}
\title{Multi-TeV vector dark matter in the $U(1)_X$ Higgs portal model}
\author[]{Kun-Peng Chen}
\author[]{Han Hu}
\author[]{Hao-Yu Luo}
\author[]{Yun Wei}
\author[]{Ming Zhong\thanks{Contact author: zhongm@nudt.edu.cn}}
\affil[]{College of Science, National University of Defense Technology, Changsha 410073, P. R. China}

\begin{document}

\maketitle

\begin{abstract}
We investigate heavy-mass vector dark matter within the framework of the $U(1)_X$ gauge Higgs portal vector model, by jointly integrating experimental constraints from relic density measurements, direct detection searches, and indirect detection observations alongside the perturbative criterion to derive bounds on the model parameter space. Within the weakly interacting massive particle scenario, we assess the relative contributions of $s$-channel and $t/u$-channel processes to thermal dark matter annihilation. For parameter configurations lying in the $s$-channel resonance region, each fixed value of the scalar mixing angle $\sin\alpha$ corresponds to a well-defined upper limit on the heavy dark scalar mass $m_{h_2}$. This upper limit decreases monotonically with increasing $\sin\alpha$, with a critical threshold of $\sin\alpha \lesssim 0.05$ beyond which perturbative constraints entirely exclude the full resonance corridor. By contrast, the non-resonant parameter regime dominated by $t/u$-channel annihilation processes is found to be excluded for $m_V \simeq 1$ TeV and above under the combined joint constraints from relic density requirements and perturbative consistency. Our results indicate that the perturbative consistency of the scalar sector, enforced via the two-loop renormalisation group equations, imposes bounds on the parameter space in the heavy mass regime that are comparable to, and in some benchmark regions more restrictive than, current experimental limits. Nevertheless, even under the application of these stringent perturbativity constraints, the model still admits a viable set of parameter points spanning a broad multi-TeV mass range.
\end{abstract}

\section{Introduction}

The Higgs portal model featuring a vector dark matter (VDM) candidate represents a simple extension of the Standard Model (SM)~\cite{Hambye:2009,Lebedev:2012,Farzan:2012,Baek:2013,Duch:2015,Ko:2014}.
In addition to the field content of the SM, this framework introduces a complex scalar singlet $S$ that obeys a $\mathrm{U}(1)_X$ gauge symmetry, alongside a gauge field $V_\mu$ corresponding to the VDM candidate. The stability of the VDM candidate is ensured by an imposed discrete $\mathbb{Z}_2$ symmetry, which concurrently forbids kinetic mixing between the dark gauge sector and the SM gauge sector. The CP-even component of the dark scalar field mixes with the SM Higgs boson, which constitutes the unique interaction portal linking the dark sector and the SM sector. This construction not only constitutes the minimal realisation of VDM, but also is capable of alleviating the vacuum stability problem of the SM and accommodating a strongly first-order electroweak phase transition that gives rise to cosmic baryon asymmetry and stochastic gravitational
waves\cite{Baek:2013,Duch:2015,Ayazi:2023,Chao:2015,Hashino:2018,Ayazi:2019}. Furthermore, the presence of two scalar mediators endows the model with richer phenomenology~\cite{Arcadi:2020,Arcadi:2024,Falkowski:2015}.

The phenomenology of this VDM model has been extensively investigated in existing academic researches~\cite{Lebedev:2012,Farzan:2012,Djouadi:2012,Baek:2013,Baek:2014,Ko:2014,Duch:2015,Beniwal:2016,Glaus:2019,Baek:2022,Amiri:2023,Ayazi:2024,Shaikh:2026}. All relevant physical observables are governed by a small set of free parameters: the VDM mass $m_V$, the dark gauge coupling $g_V$, the dark scalar mass $m_{h_2}$, and the scalar mixing angle $\alpha$. The viable parameter space is primarily constrained by multiple experimental and observational inputs, including the dark matter (DM) relic abundance, direct detection, indirect detection, and collider searches from invisible Higgs decays~\cite{Djouadi:2012,Baek:2013,Baek:2014,Duch:2015,Glaus:2019,Amiri:2023,Ayazi:2024,Beniwal:2016}. Combining all of these constraints, two distinct low-mass regions of the parameter space, spanning $m_V$ from the sub-GeV scale up to about 1 TeV, compatible with current observational requirements have been identified in the weakly interacting massive particle (WIMP) scenario. One is the $s$-channel resonant annihilation region, where a small coupling is sufficient to reproduce the observed relic abundance\cite{Shaikh:2026,Baek:2013,Duch:2015,Amiri:2023,Ko:2014}. The other is the $t/u$-channel dominated non-resonant region corresponding to a larger dark gauge coupling, where the underlying mechanism is independent of the scalar mixing angle\cite{Baek:2013,Duch:2015}. 

Nevertheless, the heavy-mass parameter regime of this model has not been systematically explored to date. In the multi-TeV mass range, all heavy SM final states, including the $WW$, $ZZ$, $t\bar{t}$, and $hh$ production channels, become kinematically accessible. Within this regime, the parameter bounds imposed by both direct and indirect DM detection experiments are substantially relaxed; concurrently, the observed DM relic abundance can be consistently reproduced without invoking finely tuned resonance configurations. Accordingly, the corresponding viable parameter space is expected to occupy a large volume, thereby warranting dedicated investigation of this parameter region in subsequent studies.

Within the heavy-mass parameter regime, increases in the masses of the VDM state and dark scalar fields yield correspondingly larger initial magnitudes of the quartic scalar couplings. As these elevated couplings undergo renormalisation group evolution, they approach the Landau pole at an accelerated rate, potentially triggering the breakdown of perturbative validity at energy scales far below the Planck scale. The analytical framework for evaluating vacuum stability and perturbativity constraints via the renormalisation group evolution of scalar couplings was systematically examined for the SM~\cite{Degrassi:2012,Buttazzo:2013}. This framework has subsequently been generalized to the VDM model~\cite{Duch:2015,Ayazi:2023}, within which a stringent perturbativity criterion has been formally established: all model couplings are required to remain below the $2\pi$ threshold across the full energy scale range extending to the Planck scale. To date, existing studies applying this framework have been largely restricted to parameter regions adjacent to the electroweak scale. Imposing two-loop renormalisation group perturbativity constraints on the TeV-scale parameter space serves as a valuable complement to evaluations of the model’s theoretical self-consistency, while also providing theoretical bounds for the heavy-mass regime that are entirely independent of experimental limits.

This study presents a targeted analysis of the VDM parameter space in the TeV-scale heavy-mass regime, integrating two-loop RGE perturbativity constraints and multi-messenger experimental bounds into a unified joint constraint framework. We combine the most recent observational datasets from direct detection, indirect detection, and relic abundance measurements with the established perturbativity criterion to constrain the parameter space associated with DM mass in the multi-TeV range, with the dark scalar mass and mixing angle fixed at representative benchmark values. Our quantitative analysis demonstrates that the model preserves a viable parameter space covering the broad DM mass interval $1\,\text{TeV} \leq m_V \leq 10\,\text{TeV}$, provided the scalar mixing angle $\alpha$ remains sufficiently small. This residual viable parameter domain is governed by the $s$-channel resonant annihilation mechanism, with the associated gauge coupling $g_V$ falling in the range from $10^{-1}$ to $10^{-2}$. Concurrently, the entire non-resonant multi-TeV regime dominated by $t/u$-channel processes is fully excluded under the combined constraints of perturbative validity requirements and cosmological relic abundance measurements.

The remainder of this paper is organized as follows. Section 2 formally introduces the VDM model, and provides explicit definitions for all model parameters relevant to the subsequent analysis. Section 3 details the implementation protocol for the experimental constraints adopted in the present work. Section 4 formally specifies the perturbativity criterion applied throughout our analysis. Section 5 presents the core results of our joint analysis, in which perturbativity constraints are integrated with experimental bounds to investigate the parameter space spanning the TeV-scale heavy-mass regime. Finally, Section 6 provides a concise summary of the key conclusions of this study.

\section{The model}
We investigate a simple extension of the SM, which introduces an additional $\mathrm{U}(1)_X$ gauged scalar theory and a scalar portal coupling to the SM Higgs ~\cite{Lebedev:2012,Farzan:2012,Duch:2015}. The dark sector of the model comprises a complex scalar field $S$ and its associated gauge boson $V_\mu$, both of which are singlets under the full SM gauge group. A discrete $\mathbb{Z}_2$ symmetry is imposed, under which the dark sector fields transform as
\begin{align}
	V_\mu \rightarrow -V_\mu, \qquad S \rightarrow S^*,
	\label{eq:Z2}
\end{align}
while all SM particles carry even parity under this $\mathbb{Z}_2$. This symmetry explicitly forbids gauge kinetic mixing between $V_\mu$ and the SM hypercharge gauge boson, thereby guaranteeing that the two gauge sectors renormalize independently at all orders in perturbation theory. As a consequence, the vector field $V_\mu$ can be identified as a stable DM candidate.

The complete Lagrangian of the model can be written as
\begin{equation}
	\mathcal{L} = \mathcal{L}_{\rm SM} - \frac{1}{4} V_{\mu\nu} V^{\mu\nu}
	+ (D_\mu S)^\dagger (D^\mu S) - V(H,S),
	\label{eq:Lagrangian}
\end{equation}
where $\mathcal{L}_{\rm SM}$ is the Lagrangian of the SM after excluding the Higgs potential term, and $V_{\mu\nu} = \partial_\mu V_\nu - \partial_\nu V_\mu$ is the field strength tensor of $V_\mu$. The covariant derivative of the scalar field $S$ is defined as
\begin{equation}
	\mathcal{D}_\mu S = (\partial_\mu + i g_V V_\mu) S,
	\label{eq:cov_deriv}
\end{equation}
where $g_V$ represents the dark gauge coupling. The most general renormalizable scalar potential that remains invariant under the gauge groups and the $\mathbb{Z}_2$ symmetry is given by
\begin{equation}
	V = -\mu_H^2 |H|^2 + \lambda_H |H|^4 - \mu_S^2 |S|^2 + \lambda_S |S|^4
	+ \lambda_{SH} |S|^2 |H|^2,
	\label{eq:potential}
\end{equation}
in which $H$ denotes the SM Higgs doublet. 

Both the electroweak and $\mathrm{U}(1)_X$ gauge symmetries undergo spontaneous breaking. In the unitary gauge, the SM Higgs field and the complex dark scalar field take the form
\begin{equation}
	H = \frac{1}{\sqrt{2}} \begin{pmatrix} 0 \\ v + \Phi_H \end{pmatrix}, \qquad
	S = \frac{1}{\sqrt{2}} (v_S + \Phi_S),
	\label{eq:VEVs}
\end{equation}
where $v \simeq 246\ \mathrm{GeV}$ and $v_S$ denote the vacuum expectation values of the SM Higgs field and the dark scalar field, respectively. The imaginary component of $S$, which corresponds to the Goldstone mode associated with the spontaneous breaking of the $\mathrm{U}(1)_X$ gauge symmetry, is absorbed by the dark photon $V_\mu$ as its longitudinal degree of freedom, thereby endowing the dark photon with a non-vanishing mass $m_V=g_V v_S$. 

The last term in Eq.~\eqref{eq:potential} gives rise to a bilinear cross term between $\Phi_H$ and $\Phi_S$, which results in a non-diagonal mass matrix of the scalars. The corresponding mass eigenstates $h_1$ and $h_2$ are obtained by diagonalising the mass matrix via the orthogonal transformation
\begin{equation}
	\begin{pmatrix} h_1 \\ h_2 \end{pmatrix}
	=
	\begin{pmatrix}
		\cos\alpha & \sin\alpha \\
		-\sin\alpha & \cos\alpha
	\end{pmatrix}
	\begin{pmatrix} \Phi_H \\ \Phi_S \end{pmatrix},
	\label{eq:mixing}
\end{equation}
where $\alpha$ denotes the scalar mixing angle. The mass eigenvalues are expressed as $m_{h_1}$ and $m_{h_2}$ respectively. In the present work, we identify $h_1$ as the experimentally observed 125 GeV SM-like Higgs boson, while $h_2$ corresponds to an additional dark scalar particle.

The coupling constants appearing in Eq.~\eqref{eq:potential} can be expressed in terms of physical quantities, as derived in~\cite{Duch:2015,Glaus:2019}
\begin{align}
	\label{eq:params1}
	\lambda_H &= \frac{m_{h_1}^2 \cos^2\alpha + m_{h_2}^2 \sin^2\alpha}{2v^2}, \\
	\label{eq:params2}
	\lambda_S &= \frac{m_{h_1}^2 \sin^2\alpha + m_{h_2}^2 \cos^2\alpha}{2v_S^2}, \\ \label{eq:params3}
	\lambda_{SH} &= \frac{(m_{h_2}^2 - m_{h_1}^2) \sin\alpha \cos\alpha}{v v_S}.
\end{align}
The requirement that the scalar potential must be bounded from below imposes the following constraints on the coupling constants
\begin{equation}
	\lambda_H > 0,\qquad \lambda_S > 0,\qquad \lambda_{SH} > -2\sqrt{\lambda_H \lambda_S}.
	\label{eq:boundedness}
\end{equation}

In summary, the model studied in this work extends the SM by introducing a spontaneously broken $\mathrm{U}(1)_X$ gauged scalar theory. It contains four additional free parameters, which are chosen as $m_{h_2}$, $m_V$, $g_V$, and $\alpha$ for notational convenience. These parameters fully govern both the physical observables that are subject to experimental constraints and the behavior of the scalar potential that is critical to the theoretical consistency of the model.

\section{Experimental constraints}

The DM phenomenology of this model has been systematically investigated in a suite of prior published studies, with existing analyses principally anchored to experimental constraints derived from DM relic density measurements \cite{Planck:2018}, direct DM detection experiments \cite{PandaX:2025,XENONnT:2025,LZ:2025}, indirect DM detection searches \cite{Fermi-LAT:2015kyq,DAMPE2022,FermiLAT:2024}, and collider searches for invisible Higgs boson decays \cite{ATLAS:2023tkt}. The physically viable parameter space of the model exhibits pronounced variation across discrete DM mass ranges.

For light DM populating the sub-GeV to GeV mass interval, the set of phenomenologically applicable constraints acting on the scenario is markedly multifaceted. Beyond the constraints outlined in the preceding discussion, the permissible parameter space is further compressed by supplementary experimental bounds, encompassing cosmic microwave background (CMB) constraints on the low-velocity s-wave annihilation cross-section, rare decay channels of light scalar states, and published measurements from beam-dump experiments~\cite{Amiri:2023}. To ensure the model retains phenomenological viability in this mass regime, two concurrent conditions must be satisfied: adequate suppression of the low-energy annihilation cross-section, achieved through either the forbidden-channel annihilation mechanism or resonant annihilation setup, to conform to CMB constraints, and implementation of an extremely small scalar mixing angle to evade collider and Higgs decay bounds. The combined effect of these superimposed requirements ultimately yields a strongly restricted overall viable parameter space.

For DM in the intermediate mass regime spanning from the GeV scale to the sub-TeV range, low-energy constraints—such as those imposed by invisible Higgs decays and rare decays of light scalar states—progressively become kinematically inaccessible, rendering the relic density requirement and direct detection bounds as the dominant constraining factors. Jointly evaluated with the relic density requirement, results from the LZ, XENONnT and PandaX experiments have excluded the overwhelming majority of non-resonant parameter configurations~\cite{LZ:2025,XENONnT:2025,PandaX:2025}, such that the remaining viable parameter space in this mass interval is predominantly concentrated in the vicinity of the dark scalar resonance. The present model can simultaneously satisfy both sets of constraints via its inherent two-scalar sector structure. For one thing, s-channel resonant annihilation mediated by the dark scalar yields a sufficiently large annihilation cross section to meet the thermal relic density requirement even in the regime of weak coupling. For another thing, destructive interference between the t-channel exchange processes mediated by the SM Higgs boson and the dark scalar serves to suppress the DM-nucleon spin-independent scattering cross section, allowing the model to evade existing direct detection limits~\cite{Baek:2013,Shaikh:2026}. Through fine-tuning of the dark scalar mass to shift the resonance peak position, the model is capable of accommodating both constraints across a relatively broad spectrum of DM masses.

As the DM mass extends into the multi-TeV heavy mass regime, the overall stringency of applied experimental constraints is substantially attenuated. One reason is that the sensitivity of direct detection experiments exhibits a progressive decline as DM mass increases. Another is that low-energy constraints, including those from invisible Higgs decays and rare decays of light scalar states, cease to be applicable due to the closure of their corresponding kinematic thresholds. Nevertheless, systematic phenomenological investigations targeting this regime remain comparatively sparse. Furthermore, the majority of existing studies are conducted within the scale-invariant radiative symmetry breaking framework, in which model parameters are correlated via radiative corrections and thus cannot be directly generalized to the standard UV-complete model with tree-level mass terms~\cite{Frandsen:2023}. For the latter case, the phenomenological behavior and full viable parameter space in the heavy mass regime have yet to receive systematic characterization.

\subsection{Relic density}
The relic abundance of DM is determined by its thermal history in the early Universe~\cite{Kolb:1990}. The present cold DM density has been measured with high precision by the Planck satellite, yielding $\Omega_{\mathrm{DM}} h^2 = 0.1200 \pm 0.0012$~\cite{Planck:2018}. Within the standard thermal freeze-out framework, the evolution of the DM number density $n$ is governed by the Boltzmann equation
\begin{equation}
	\frac{dn}{dt} + 3H n = -\langle\sigma v\rangle (n^2 - n_{\rm eq}^2),
	\label{eq:Boltzmann}
\end{equation}
where $H$ represents the Hubble expansion rate, $n_{\rm eq}$ denotes the equilibrium number density, and $\langle\sigma v\rangle$ corresponds to the thermally averaged annihilation cross section.

\begin{figure}[t]
    \centering
    \begin{subfigure}[b]{0.30\textwidth}
        \centering
        \begin{tikzpicture}[baseline=(current bounding box.center)]
            \begin{feynman}
                \vertex (i1) at (0,1) {$V$};
                \vertex (i2) at (0,-1) {$V$};
                \vertex (a) at (1,0);
                \vertex (b) at (2,0);
                \vertex (f1) at (3,1) {$\mathrm{SM}$};
                \vertex (f2) at (3,-1) {$\mathrm{SM}$};
                \diagram*{
                    (i1) -- [boson] (a) -- [boson] (i2),
                    (a) -- [scalar, edge label=$h_{1,2}$] (b),
                    (f1) -- [plain] (b) -- [plain] (f2),
                };
            \end{feynman}
        \end{tikzpicture}
    \end{subfigure}
    \hfill
    \begin{subfigure}[b]{0.30\textwidth}
        \centering
        \begin{tikzpicture}[baseline=(current bounding box.center)]
            \begin{feynman}
                \vertex (i1) at (0,1) {$V$};
                \vertex (i2) at (0,-1) {$V$};
                \vertex (a) at (1,1);
                \vertex (b) at (1,-1);
                \vertex (f1) at (3,1) {$h_{1,2}$};
                \vertex (f2) at (3,-1) {$h_{1,2}$};
                \diagram*{
                    (i1) -- [boson] (a) -- [scalar] (f1),
                    (i2) -- [boson] (b) -- [scalar] (f2),
                    (a) -- [boson, edge label=$V$] (b),
                };
            \end{feynman}
        \end{tikzpicture}
    \end{subfigure}
    \hfill
    \begin{subfigure}[b]{0.30\textwidth}
        \centering
        \begin{tikzpicture}[baseline=(current bounding box.center)]
            \begin{feynman}
                \vertex (i1) at (0,1) {$V$};
                \vertex (i2) at (0,-1) {$V$};
                \vertex (a) at (1.5,0);
                \vertex (f1) at (3,1) {$h_{1,2}$};
                \vertex (f2) at (3,-1) {$h_{1,2}$};
                \diagram*{
                    (i1) -- [boson] (a) -- [boson] (i2),
                    (f1) -- [scalar] (a) -- [scalar] (f2),
                };
            \end{feynman}
        \end{tikzpicture}
    \end{subfigure}
   \caption{Tree-level Feynman diagrams for DM annihilation processes in the VDM model.
   	Left panel: $s$-channel annihilation into SM final states proceeding via $h_1$ or $h_2$ exchange.
   	Middle panel: $t$- and $u$-channel annihilation into scalar pairs.
   	Right panel: annihilation via the quartic contact interaction.}
    \label{fig:feynman}
\end{figure}
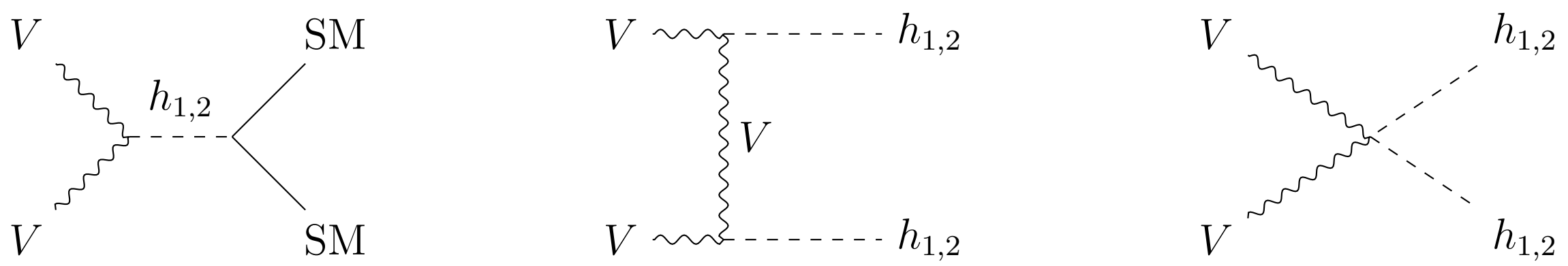

As illustrated in Figure~\ref{fig:feynman}, the dominant DM annihilation channels of the present model consist of two topological classes: $s$-channel annihilation transitions mediated via $h_1$ and $h_2$ scalar exchange that produce SM final states, and the $VV \to h_{1,2}h_{1,2}$ process arising from the combined contributions of $t/u$-channel topologies and direct contact coupling at the scalar quartic interaction vertex. For the subsequent evaluation of phenomenological constraints applicable to the investigated heavy DM mass regime, the cross-section contributions originating from these two process classes are analyzed in isolation. This discrete analytical decomposition is underpinned by robust physical justification. The two process classes exhibit intrinsic disparities in their interaction vertices and their parametric dependencies, and accordingly contribute with markedly distinct magnitudes to the dark scalar resonant and non-resonant regions of the parameter space.

The annihilation cross section corresponding to the $s$-channel scalar exchange process follows the scaling form $g_V^2 \sin^22\alpha$. As the center-of-mass energy nears the mass of either scalar eigenstate, the cross section manifests marked resonant enhancement induced by the pole structure of the associated propagator. Given that this process exhibits an identical dependence on interaction couplings and mixing angle to that of the DM–nucleon scattering cross section, resonant enhancement constitutes the central mechanism enabling simultaneous compliance with the thermal relic density requirement and direct detection experimental constraints in the regime of small gauge couplings.

Conversely, the annihilation cross section arising from $t/u$-channel vector exchange and the quartic contact vertex follows the scaling behavior $g_V^4 \cos^4\alpha$. These non-resonant processes constitute the dominant contribution to the total annihilation cross section in parameter regimes well separated from the scalar resonance poles, particularly in the large gauge coupling domain. The mixing angle dependence of these processes is non-identical to that of the direct detection process. Consequently, the direct detection cross section may be suppressed via reduction of the mixing angle, while retaining an annihilation cross section of sufficient magnitude to meet the relic density requirement, thus realizing partial decoupling of the two constraints.

The model under study is implemented in \texttt{MicrOMEGAs}~6.0~\cite{MicrOMEGAs:2023} through the dedicated interface of \texttt{SARAH}~\cite{SARAH4,SARAH3,SARAH:Unitarity,SARAH:Goldstone} and \texttt{SPheno}~\cite{SPheno:2003,SPheno:2012}, with these packages collectively generating the full particle mass spectrum, interaction couplings, and complete set of Feynman rules required for phenomenological calculations. \texttt{MicrOMEGAs} subsequently computes the thermally averaged annihilation cross section for all kinematically open channels, and performs a numerical solution of Eq.~\eqref{eq:Boltzmann} to derive the present-day DM relic abundance. In the analysis presented here, we impose the constraint that the theoretically predicted relic density does not exceed the observationally measured value, thereby allowing the VDM candidate to account for only a fraction of the total cosmological DM density.

\subsection{Direct detection}

Direct detection experiments are dedicated to detecting nuclear recoils induced by the scattering of Galactic DM particles off target nuclei installed in deep underground laboratories~\cite{Schumann:2019}. Currently, the most stringent upper limits on the spin-independent (SI) DM-nucleon scattering cross section are established by the LUX-ZEPLIN (LZ) experiment~\cite{LZ:2025}, with complementary constraints provided by the XENONnT and PandaX experiments~\cite{XENONnT:2025,PandaX:2025}.

In the VDM model, the SI scattering of VDM off a nucleon proceeds via $t$-channel exchange of the two scalar mediators $h_1$ and $h_2$. At tree level, the resulting scattering cross section is formulated as~\cite{Glaus:2019}

\begin{equation}
	\sigma^{\rm SI}_N = \frac{\mu_N^2}{4\pi}\,
	g_V^2\, \sin^2 2\alpha\,
	\left(\frac{1}{m_{h_1}^2} - \frac{1}{m_{h_2}^2}\right)^{\!2}
	\frac{m_N^2}{v^2}\, f_N^2,
	\label{eq:DD}
\end{equation}
where $\mu_N = m_V m_N/(m_V + m_N)$ denotes the reduced mass of the DM-nucleon system, and $f_N$ is the effective Higgs--nucleon coupling. A distinctive feature of this model is the destructive interference between the $h_1$- and $h_2$-mediated exchange amplitudes, which is encoded in the factor $(m_{h_1}^{-2}-m_{h_2}^{-2})^2$. This effect can significantly suppress the scattering cross section when the two scalars are nearly mass-degenerate~\cite{Baek:2013,Duch:2015}.

In the present analysis, we adopt the latest 2025 LZ exclusion limit on $\sigma^{\rm SI}_N$. Given that the effective local DM density is attenuated when the VDM candidate accounts for only a fraction $f = \Omega/\Omega_{\rm obs}$ of the total observationally measured relic abundance, we rescale the published experimental bound to obtain the effective cross section $\sigma^{\rm eff} = f \cdot \sigma^{\rm SI}_N$, in strict accordance with established standard analytical protocols.

\subsection{Other experimental constraints}
Two additional experimental constraints, which have been extensively employed in existing literature, are found to be of very limited relevance for the heavy mass region investigated in this work. The invisible decay of the SM-like Higgs boson, $h_1\to VV$, yields a stringent constraint when $m_V<m_{h_1}/2$~\cite{Lebedev:2012,Djouadi:2012}, yet this decay process is kinematically forbidden for $m_V>62.5$~GeV. For indirect detection constraints, we follow the approach of Ref.~\cite{Glaus:2019} and adopt the Fermi-LAT $b\bar{b}$-channel upper limit from the 14-year dSph legacy analysis~\cite{FermiLAT:2024} as a representative bound on the annihilation cross section. We calculate the velocity-averaged total annihilation cross section $\langle\sigma v\rangle_{\rm tot}$ using \texttt{MicrOMEGAs}, and compare the effective cross section $\langle\sigma v\rangle_{\rm eff}=f^{2}\langle\sigma v\rangle_{\rm tot}$ with the $b\bar{b}$-channel upper limit, where $f^2$ is the required rescaling factor for a subdominant DM component. Across the full heavy mass regime, the constraints imposed by indirect detection are observed to be substantially less stringent than those arising from the measured relic abundance and direct detection. Consequently, the subsequent analysis presented in this work is governed dominantly by constraints originating from relic abundance, direct detection, and perturbativity. 

\section{The renormalisation group equations and perturbativity constraints}

At the electroweak scale, the quartic scalar couplings of the VDM model can be selected to lie within the perturbative regime while satisfying all existing experimental constraints. However, the heavy dark sector mass regime requires larger magnitudes of the dark gauge coupling $g_V$ and the dark vacuum expectation value $v_S$, which in turn drives the initial values of the scalar couplings to deviate from their electroweak-scale benchmark configurations. Under renormalisation group evolution, these enhanced couplings may approach a Landau pole at energy scales far below the Planck scale, rendering the theory non-perturbative. Consequently, any parameter point that appears phenomenologically viable at low energy scales can be excluded once renormalisation group running effects are incorporated, in accordance with the imposed perturbativity constraints.

The renormalization group equations (RGEs) describing the scale evolution of all model parameters for the VDM are derived with the \texttt{SARAH} package~\cite{SARAH4,SARAH3,SARAH:Unitarity,SARAH:Goldstone}, which yields the complete set of two-loop-order $\beta$ functions employed for all numerical calculations reported in this work. In line with the methodology established in Ref.~\cite{Duch:2015}, the boundary initial conditions for the renormalization group running are fixed at the top-quark mass scale $\mu=m_t$, with all coupling parameters numerically evolved along the renormalization scale up to the Planck scale $M_{\rm Pl}\simeq 1.22\times10^{19}$~GeV. For illustrative clarity, we present the one-loop-order $\beta$ functions for the dark gauge coupling $g_V$ and the scalar quartic couplings
\begin{align}
	(16\pi^2)\beta_{g_V}^{(1)} &= \frac13\,g_V^{3},
	\label{eq:beta_gV}\\[4pt]
	(16\pi^2)\beta_{\lambda_H}^{(1)} &=
	\frac38 g_1^4 + \frac34 g_1^2 g_2^2 + \frac98 g_2^4
	-3g_1^2\lambda_H -9g_2^2\lambda_H
	+24\lambda_H^2 + \lambda_{SH}^2 \nonumber\\
	&\quad +12\lambda_H\,{\rm Tr}\big(Y_d Y_d^\dagger\big)
	+4\lambda_H\,{\rm Tr}\big(Y_e Y_e^\dagger\big)
	+12\lambda_H\,{\rm Tr}\big(Y_u Y_u^\dagger\big) \nonumber\\
	&\quad -6\,{\rm Tr}\big(Y_d Y_d^\dagger Y_d Y_d^\dagger\big)
	-2\,{\rm Tr}\big(Y_e Y_e^\dagger Y_e Y_e^\dagger\big)
	-6\,{\rm Tr}\big(Y_u Y_u^\dagger Y_u Y_u^\dagger\big),
	\label{eq:beta_lambdaH}\\[4pt]
	(16\pi^2)\beta_{\lambda_{SH}}^{(1)} &=
	\frac{\lambda_{SH}}{2}\Bigl(-3g_1^2-9g_2^2-12g_V^2
	+24\lambda_H+8\lambda_{SH}+16\lambda_S \nonumber\\
	&\quad +12\,{\rm Tr}\big(Y_d Y_d^\dagger\big)
	+4\,{\rm Tr}\big(Y_e Y_e^\dagger\big)
	+12\,{\rm Tr}\big(Y_u Y_u^\dagger\big)\Bigr),
	\label{eq:beta_lambdaSH}\\[4pt]
	(16\pi^2)\beta_{\lambda_S}^{(1)} &=
	2\bigl(3g_V^{4}+\lambda_{SH}^{2}-6g_V^{2}\lambda_S+10\lambda_S^{2}\bigr).
	\label{eq:beta_lambdaS}
\end{align}
Within the foregoing expressions, $g_1$, $g_2$, and $g_3$ denote the SM gauge couplings, while $Y_{u,d,e}$ represent the Yukawa coupling matrices. The one-loop-order $\beta$ functions corresponding to the three aforementioned SM gauge couplings are given by
\begin{equation}
	(16\pi^2)\beta_{g_1}^{(1)} = \frac{41}{6}\,g_1^{3},\qquad
	(16\pi^2)\beta_{g_2}^{(1)} = -\frac{19}{6}\,g_2^{3},\qquad
	(16\pi^2)\beta_{g_3}^{(1)} = -7\,g_3^{3}.
\end{equation}

Throughout the present work, we implement the perturbativity requirement stipulating that all model coupling parameters remain bounded below $2\pi$ across the full renormalization scale range up to $M_{\rm Pl}$~\cite{Duch:2015}. Any parameter point for which at least one coupling exceeds this threshold is identified as hosting a Landau pole, and is consequently excluded from the set of viable parameter configurations. The one-loop-order beta function $\beta_{g_V}^{(1)}$ is positive, indicating that the renormalization group evolution of $g_V$ proceeds in a direction aligned with the variation of the renormalization energy scale $\mu$. It should be noted that for an initial dark gauge coupling \(g_{V,0} = 1\), the scale associated with the onset of the corresponding Landau pole is of order \(\boldsymbol{\sim 10^{105}\,\text{GeV}}\), a value substantially larger than the Planck scale. As a direct consequence, the running coupling \(g_V\) imposes no perturbativity constraint across the full parameter coverage of the subsequent analysis. The signs of the remaining one-loop beta functions $\beta_{\lambda_H}^{(1)}$, $\beta_{\lambda_{SH}}^{(1)}$ and $\beta_{\lambda_S}^{(1)}$ are contingent on the specific numerical values of the model’s intrinsic coupling constants, which introduces additional analytical complexity to the present study.

\section{Parameter space of the heavy-mass region}

Investigation of the heavy-mass parameter space is conducted in the present analysis. The dark scalar mass $m_{h_2}$ and the mixing angle $\sin\alpha$ are fixed at representative benchmark values, while the VDM mass $m_V$ and dark gauge coupling $g_V$ are varied over dedicated scanning domains. For each benchmark configuration, $m_V$ is sampled across a parameter range centered on the resonance condition $m_V \approx m_{h_2}/2$, and $g_V$ is sampled over the bounded interval $[10^{-4}, \sqrt{4\pi}]$. For every sampled parameter point, the thermal relic density is calculated with the \texttt{MicrOMEGAs} package, whereas the spin-independent DM-nucleon direct detection cross section is evaluated using the closed-form analytical expression presented in Eq.~\eqref{eq:DD}, with results benchmarked against the most recent LZ experimental exclusion limit. The two-loop-order perturbativity criterion derived via the \texttt{SARAH} toolchain is applied to the full set of sampled points. Only parameter points that satisfy all three imposed constraints are retained as members of the physically viable parameter space.

\begin{figure}[t]
	\centering
	\begin{minipage}[t]{0.48\textwidth}
		\centering
		\includegraphics[width=\textwidth,height=6cm]{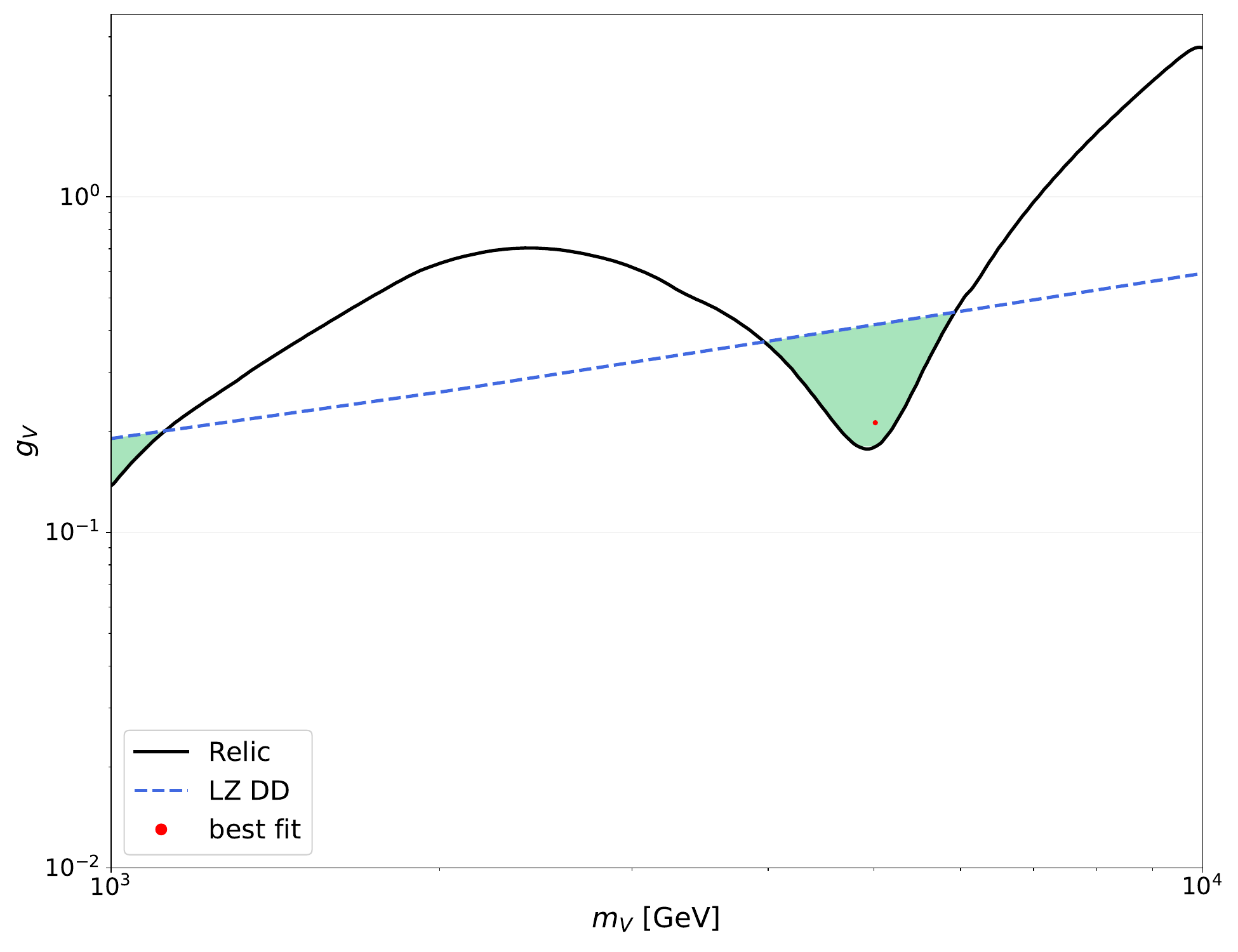}
	\end{minipage}
	\hfill
	\begin{minipage}[t]{0.48\textwidth}
		\centering
		\includegraphics[width=\textwidth,height=6cm]{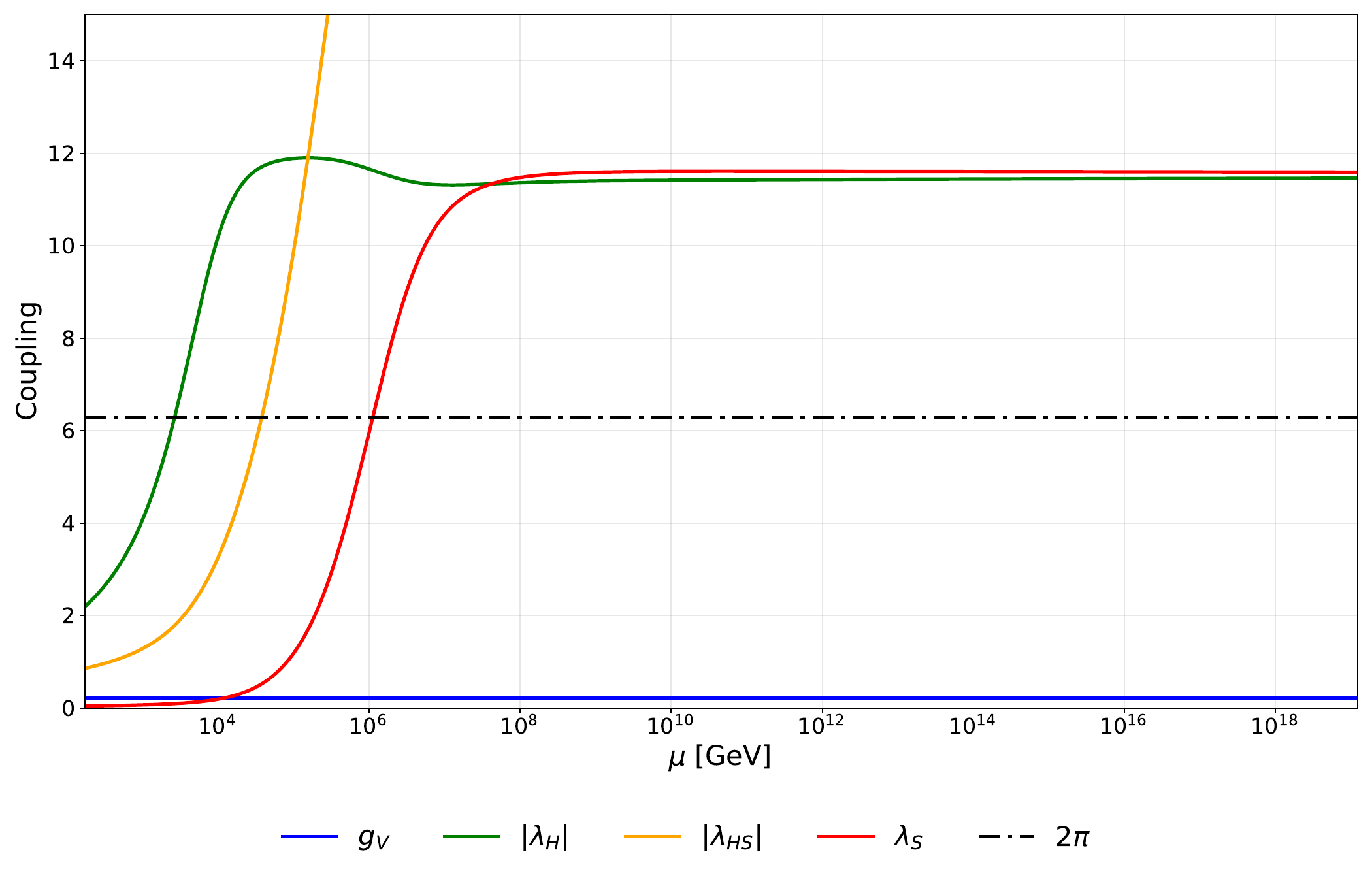}
	\end{minipage}
	\caption{A representative illustration of identifying the physically admissible parameter domains. Left panel: Superposition of integrated experimental constraints on the $(m_V,g_V)$ parameter plane for the specified input values $m_{h_2}=10$~TeV and $\sin\alpha=0.05$. Right panel: RGE evolution of the scalar quartic couplings for the benchmark parameter point defined by $g_V=0.212$ and $m_V=5$~TeV.}
	\label{fig:10TeV}
\end{figure}

A representative exemplary case is provided below to demonstrate the previously elaborated analytical workflow, with dedicated focus on the implementation mechanism of the perturbativity criterion during the identification of physically allowed parameter domains. For the specified benchmark scenario defined by $m_{h_2}=10\ \text{TeV}$ and $\sin\alpha=0.05$, the resonant parameter domain of $m_V$ is nominally centered at the scale of $5\ \text{TeV}$. The left panel of Figure~\ref{fig:10TeV} visualizes the $(m_V, g_V)$ parameter domain restricted by all currently available experimental observations, in which the overlaid red marker identifies the global best-fit configuration. The right panel presents the renormalization group evolution trajectories of the associated scalar quartic couplings. For illustrative purposes, we select a constraint-satisfying parameter point $(g_V,m_V)=(0.212,5\,\text{TeV})$ lying within the allowed region shown in the left panel as the test case. The initial boundary values of the model couplings associated with this test point are derived directly from Eqs.~(\ref{eq:params1})--(\ref{eq:params3}). With these initial boundary conditions specified, we solve the renormalization group equation $\frac{d\lambda}{d\ln\mu}=\beta$ to obtain the running profiles of the three quartic couplings $\lambda_H$, $\lambda_{HS}$ and $\lambda_S$, all of which exhibit monotonically increasing behavior as the renormalization energy scale $\mu$ rises. The running value of $\lambda_H$ first surpasses the upper perturbativity bound $2\pi$ at the scale $Q\sim 10^{3}$~GeV, while $\lambda_{HS}$ and $\lambda_S$ sequentially exceed the same threshold at progressively higher energy scales. This observation provides explicit proof that perturbativity is violated at an energy scale well below the Planck scale. Despite the full consistency of this parameter point with all low-energy experimental constraints, it is formally excluded from the set of viable parameter configurations by the renormalization group running-based perturbativity criterion.

\subsection{The s-channel resonance region}
\begin{figure*}[!t]
  \centering
  \includegraphics[width=\textwidth]{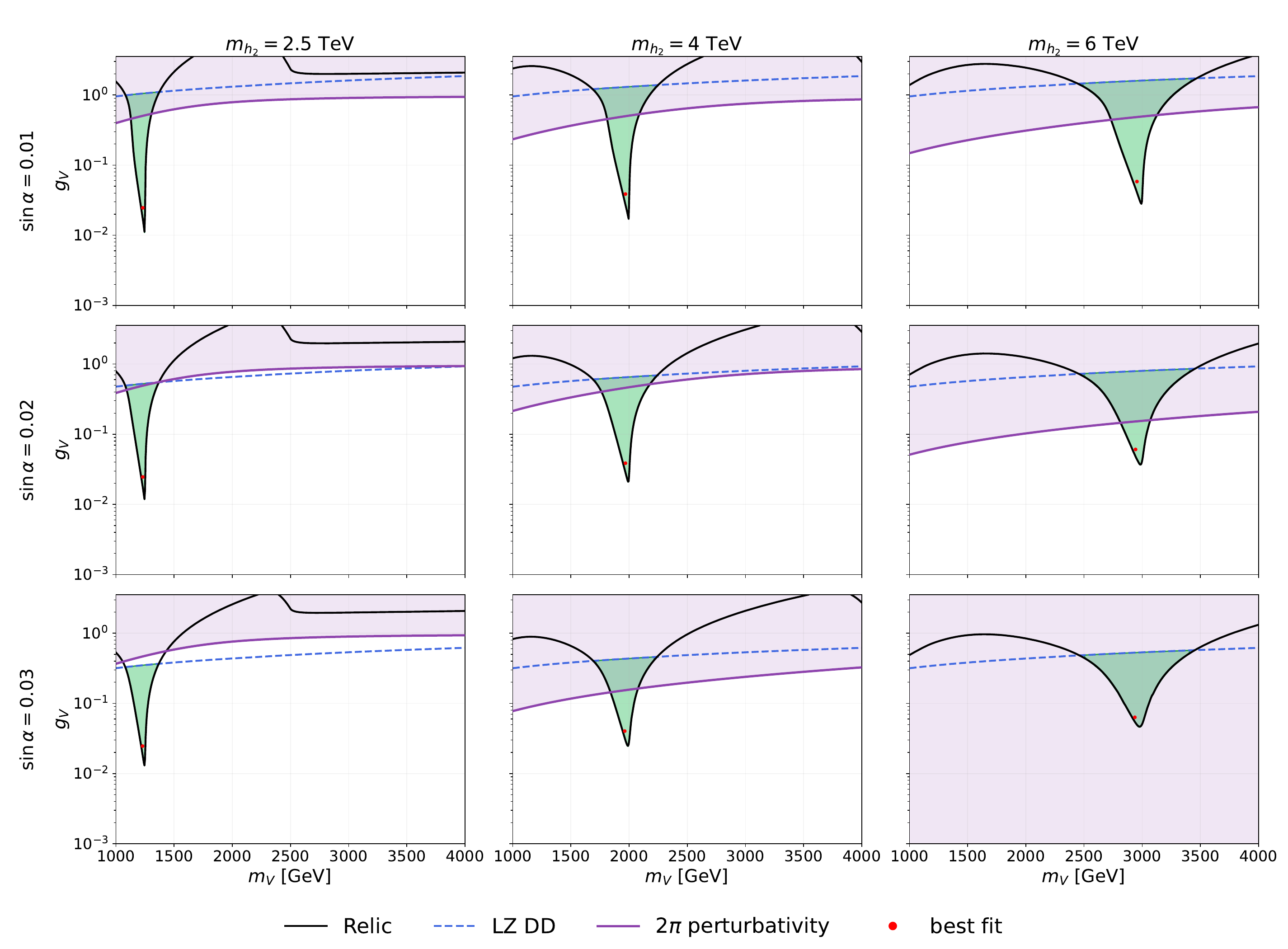}
  \caption{Parametric domain projected onto the $(m_V, g_V)$ plane. The green-shaded continuous band denotes the experimentally permitted parameter regions evaluated at the $3\sigma$ confidence level. The solid purple curve delineates the $2\pi$ perturbativity boundary, where the parameter domain excluded for points above this boundary is rendered with hatched shading. The red circular marker embedded in each panel locates the corresponding best-fit parameter point.
  }
  \label{fig:grid3x3}
\end{figure*}

To characterize the heavy-mass parameter regime accessible to s-channel resonant processes, we construct a suite of discretized benchmark parameter configurations specified by $\sin\alpha = 0.01,\,0.02,\,0.03$ and $m_{h_2} = 2.5,\,4,\,6$~TeV. For each of the nine benchmark parameter combinations, we carry out a grid-based parameter scan: the $m_V$ dimension is sampled via uniform linear discretization, while the $g_V$ dimension is sampled via uniform logarithmic discretization, with approximately $10^5$ independent numerical evaluation points deployed per configuration. The relic abundance contour satisfying $\Omega h^2 = 0.12$ and the $2\pi$ perturbativity boundary are independently resolved using iterative numerical bisection. The resulting allowed parameter domains, projected onto the two-dimensional $(m_V,\,g_V)$ plane, are displayed in Fig.~\ref{fig:grid3x3}. The panel layout is structured such that panels progress sequentially along the horizontal axis for $m_{h_2}=2.5$, $4$ and $6\ \mathrm{TeV}$, and along the vertical axis for $\sin\alpha=0.01$, $0.02$ and $0.03$, arranged in descending order from top to bottom. The green shaded regions denote the allowed parameter domains subject to joint constraints from measured dark matter relic density, LZ direct detection searches, and Fermi-LAT indirect detection observations, evaluated at the $3\sigma$ confidence level. The solid black curve traces the relic abundance contour, and the dashed blue curve corresponds to the LZ direct-detection exclusion limit. The solid purple curve marks the $2\pi$ perturbativity boundary, with the parameter region excluded above this boundary rendered as shaded hatching. The extracted numerical parameter ranges for these nine benchmark configurations, which present a direct comparison between the bounds derived from standalone experimental constraints and those obtained under the combined experimental and perturbativity constraints, are compiled in Table~\ref{tab:paramspace}.
\begin{table}[htbp]
  \centering
  \caption{Numerical results of the allowed parameter space for the nine benchmark points, derived separately under two distinct constraint sets: standalone experimental evaluations conducted at a \(3\sigma\) confidence level, and the joint combined implementation of experimental bounds and perturbative consistency constraints.}
  \label{tab:paramspace}
  \small 
  \setlength{\tabcolsep}{0pt} 
  \begin{tabular}{ c @{\hspace{12pt}} c @{\hspace{20pt}} c @{\hspace{8pt}} c @{\hspace{20pt}} c @{\hspace{8pt}} c }
    \toprule
    & &
    \multicolumn{2}{c}{Experimental only} &
    \multicolumn{2}{c}{\shortstack{Experimental $+$ \\ perturbativity}} \\ 
    \cmidrule(lr){3-4}\cmidrule(lr){5-6}
    \(m_{h_2}\) [TeV] & \(\sin\alpha\) & \(m_V\) [GeV] & \(g_V\) & \(m_V\) [GeV] & \(g_V\) \\
    \midrule
    \multirow{3}{*}{2.5}
      & 0.01 & 1094\,--\,1359 & 0.010\,--\,1.077 & 1129\,--\,1305 & 0.010\,--\,0.536 \\
      & 0.02 & 1090\,--\,1357 & 0.012\,--\,0.536 & 1094\,--\,1357 & 0.012\,--\,0.536 \\
      & 0.03 & 1090\,--\,1355 & 0.013\,--\,0.355 & 1090\,--\,1355 & 0.013\,--\,0.355 \\
    \midrule
    \multirow{3}{*}{4}
      & 0.01 & 1710\,--\,2226 & 0.017\,--\,1.323 & 1812\,--\,2085 & 0.017\,--\,0.514 \\
      & 0.02 & 1712\,--\,2220 & 0.021\,--\,0.686 & 1762\,--\,2180 & 0.021\,--\,0.514 \\
      & 0.03 & 1714\,--\,2218 & 0.025\,--\,0.455 & 1840\,--\,2077 & 0.025\,--\,0.163 \\
    \midrule
    \multirow{3}{*}{6}
      & 0.01 & 2492\,--\,3420 & 0.028\,--\,1.693 & 2732\,--\,3129 & 0.028\,--\,0.514 \\
      & 0.02 & 2498\,--\,3411 & 0.037\,--\,0.842 & 2814\,--\,3069 & 0.037\,--\,0.156 \\
      & 0.03 & 2502\,--\,3408 & 0.048\,--\,0.558 & \multicolumn{2}{c}{---} \\
    \bottomrule
  \end{tabular}
\end{table}

Across all nine sampled ($m_{h_2}, \sin\alpha$) parameter combinations, experimentally admissible regions within the ($m_V, g_V$) parameter space are identified. Holding $\sin\alpha$ constant, the lower limit on $g_V$ extracted from cosmological DM relic density constraints exhibits a monotonic increasing trend with rising $m_{h_2}$, whereas the upper limit on $g_V$ set by LZ direct detection measurements remains nearly insensitive to variations in $m_{h_2}$. When $m_{h_2}$ is held fixed, the relic-density-derived lower limit on $g_V$ rises as $\sin\alpha$ increases, while the LZ direct-detection upper limit on $g_V$ displays a systematic decreasing trend with growing $\sin\alpha$. As a direct consequence, the resonant parameter region jointly bounded by relic density constraints and LZ direct detection limits expands as $m_{h_2}$ increases or $\sin\alpha$ decreases, accompanied by a synchronous upward shift of the minimum allowed value of $g_V$. With increasing $m_{h_2}$, this resonant region migrates progressively toward higher $m_V$ values, while its centroid remains marginally positioned below the on-shell resonance pole $m_{h_2}/2$. 

The above observed phenomenological behaviour can be qualitatively interpreted by referencing the intrinsic decay width of the dark scalar. The total decay width of the $h_2$ state scales positively with both the mixing angle parameter $\sin\alpha$ and the dark scalar mass $m_{h_2}$. An increment in $m_{h_2}$ consequently yields a wider resonance profile with a correspondingly reduced resonant peak height. Given that $h_2$ decays dominantly into SM final states, the resonant DM annihilation cross section is effectively diluted. As a result, a larger gauge coupling $g_V$ is mandated to recover the observed relic abundance condition $\Omega h^2=0.12$. The resonant condition is fundamentally located at center-of-mass energy $\sqrt{s}\approx m_{h_2}$, while the center-of-mass energy of the DM pair is only $\sqrt{s}\approx2m_V$, with the energy offset bridging this gap provided by the thermal kinetic energy of the DM particles. As the freeze-out temperature $T_f$ scales proportionally with $m_{h_2}$, an increase in $m_{h_2}$ elevates the mean thermal kinetic energy of the DM population. Correspondingly, the thermal energy spread of the DM ensemble also grows with rising $T_f$, such that parameter points further from the on-shell resonant condition $m_V=m_{h_2}/2$ retain sufficient kinetic energy to satisfy the resonant annihilation requirement, thereby broadening the allowed $m_V$ parameter window.

Upon the incorporation of the perturbativity constraint, most of the experimentally allowed parameter regions across the nine benchmark parameter combinations are subjected to significantly more stringent restrictions. Specifically, for the benchmark point defined by $m_{h_2}=6\ \mathrm{TeV}$ and $\sin\alpha=0.03$, the entire spanned parameter space is rendered fully excluded. At fixed $\sin\alpha$, the perturbativity-derived upper limit on $g_V$ exhibits a monotonic decreasing trend with rising $m_{h_2}$, corresponding to a progressively tighter restrictive bound. At fixed $m_{h_2}$, the perturbativity-derived upper limit on $g_V$ likewise decreases monotonically with increasing $\sin\alpha$. As a direct consequence, for each fixed $\sin\alpha$ value, the perturbativity condition defines a hard upper cutoff on $m_{h_2}$. Evaluating this cutoff under the on-shell resonance condition $m_V=m_{h_2}/2$, we derive the corresponding maximum permitted mass values $m_{h_2}^{\max}=12.15$, $6.08$ and $4.05\ \mathrm{TeV}$ for $\sin\alpha=0.01$, $0.02$ and $0.03$, respectively. At fixed $\sin\alpha$, as $m_{h_2}$ approaches this upper cutoff, the fraction of allowed parameter space eliminated by the perturbativity constraint grows progressively, until the cutoff is reached at which the entire parameter region becomes excluded. Consistent with this trend, the full parameter space associated with $m_{h_2}=6\ \mathrm{TeV}$ (which exceeds the $4.05\ \mathrm{TeV}$ upper cutoff for $\sin\alpha=0.03$) is fully excluded by the perturbativity requirement. An additional relevant finding is that, to ensure the resonant annihilation valley remains located within the TeV-scale regime, $\sin\alpha$ must not exceed 0.05, which corresponds to a maximum permitted heavy scalar mass $m_{h_2}^{\max}=2.43\ \mathrm{TeV}$.

In summary, upon imposition of the perturbativity constraint, for each fixed value of the mixing angle $\sin\alpha$, a well-defined upper limit $m_{h_2}^{\max}$ on the dark scalar mass is identified. For all dark scalar masses satisfying $m_{h_2}<m_{h_2}^{\max}$, a resonant valley emerges in the parameter region $m_V\simeq m_{h_2}/2$; the fraction of this valley excluded by the perturbativity requirement grows progressively as $m_{h_2}$ approaches $m_{h_2}^{\max}$, until the entire parameter range of the valley is fully excluded when $m_{h_2}$ reaches its upper bound. The nontrivial character of the large dark scalar mass regime arises directly from the dependence of the resonance width on $m_{h_2}$, as this quantity modulates the depth of the resonant valley. Extrapolation of the derived parameter scaling indicates that smaller mixing angle values, which yield $m_{h_2}^{\max}>20\ \mathrm{TeV}$, permit DM mass solutions that satisfy both experimental constraints and perturbativity requirements across an extended multi-TeV interval. This analysis demonstrates that, even when the stringent perturbativity requirement is enforced, the admissible DM mass range of the considered model spans the broad interval from $1$~TeV to $10$~TeV.

\subsection{The t/u-channel dominated non-resonant region}

\begin{figure}[t]
    \centering
    \includegraphics[width=0.75\textwidth]{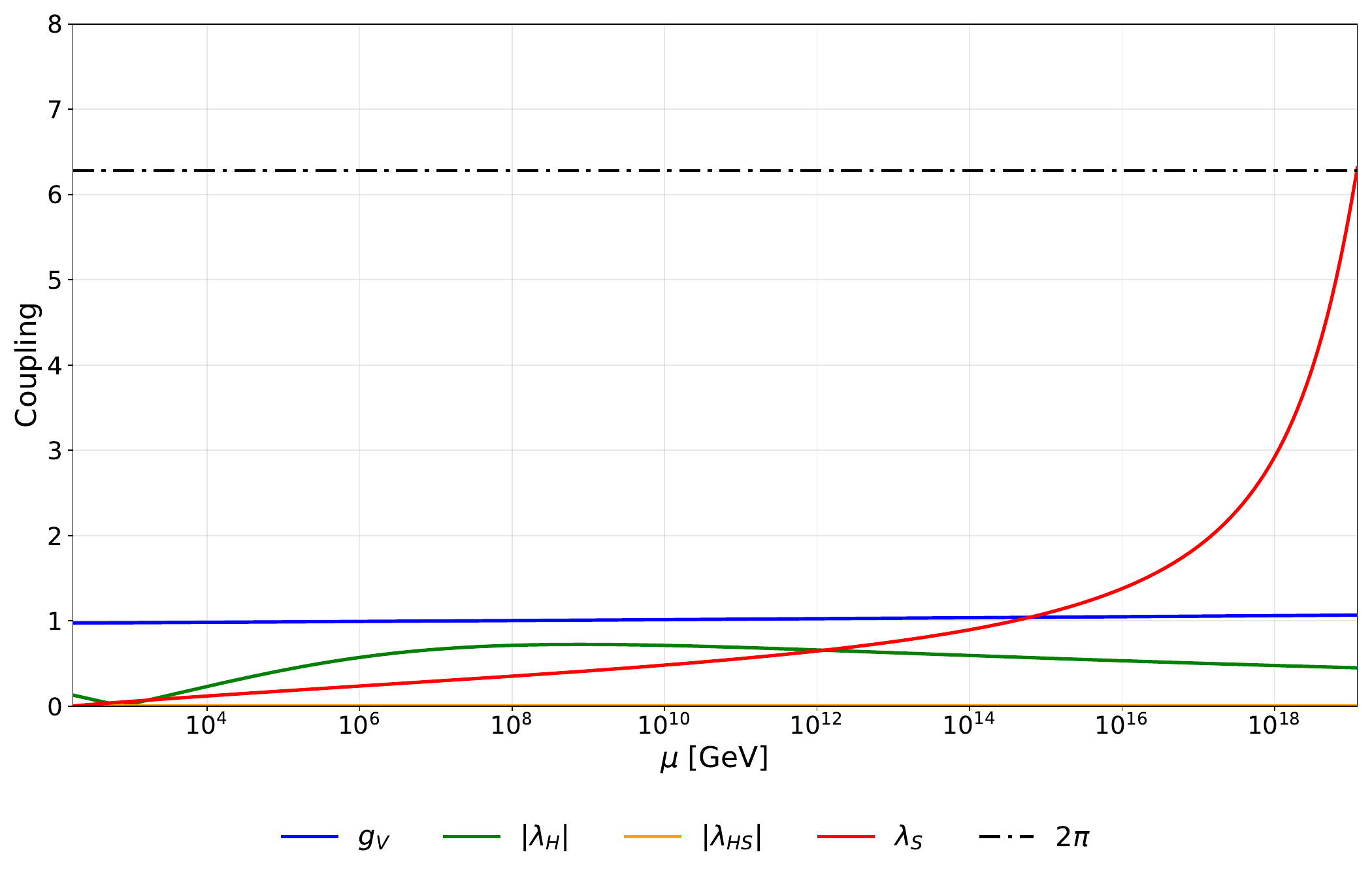}
    \caption{Two-loop RGE evolution of scalar quartic couplings for the benchmark point characterized by $m_{h_2}=100$~GeV, $\sin\alpha=0.01$, $m_V=1$~TeV, and $g_V=0.974$. }
    \label{fig:rge_critical}
\end{figure}

For configurations with a comparatively light dark scalar satisfying $m_{h_2}\lesssim m_V$, the leading DM annihilation pathway falls into the non-resonant regime, where dominant contributions arise from the $t$-channel and $u$-channel exchange processes alongside the four-point contact interaction. The thermally averaged annihilation cross section associated with these processes obeys the scaling relation $\langle\sigma v\rangle_{t}\propto g_V^{4}\cos^{4}\alpha$. For the low-mass regime, this scaling permits selection of a sufficiently small $\sin\alpha$ value to evade current direct detection experimental constraints, while the $t$-channel exchange contribution remains sufficient large to reproduce the observed DM relic abundance. By contrast, in the high-mass regime, the allowed range of $g_V$ is subject to stringent restrictions imposed by the perturbativity requirement. To systematically examine this parameter regime in isolation, we adopt a benchmark scenario defined by a light dark scalar mass $m_{h_2}=100\ \mathrm{GeV}$ and a small mixing angle $\sin\alpha=0.01$, a parameter choice that minimizes the s-channel annihilation contribution and strongly suppresses the direct detection scattering cross section. The maximum gauge coupling permitted by the two-loop RGEs is $g_V^{\rm max}=0.974$. The corresponding Two-loop RGE running trajectory is presented in Fig.~\ref{fig:rge_critical} where the coupling $\lambda_S$ attains the perturbativity threshold of $2\pi$ at the Planck mass scale $M_{\rm Pl}$.

We further examine whether the perturbativity bound on $g_V$ can be alleviated by varying the mixing angle $\sin\alpha$ or by increasing $m_V$. Neither adjustment has any appreciable effect: across all tested configurations the derived $g_V^{\rm max}$ remains close to $0.974$. This behavior originates from the structural form of the renormalization group beta function for $\lambda_S$: the one-loop contribution includes a term proportional to $g_V^{4}$, whereas the two-loop correction scales with $g_V^{6}$. While a reduced initial value of $\lambda_S$ retards the emergence of the Landau pole, the rapid running of $\lambda_S$ induced by the large gauge coupling $g_V$ offsets this delaying effect, leading to perturbativity violation at an almost identical energy scale that is largely insensitive to the chosen input parameters. Accordingly, we conclude that $g_V=0.974$ constitutes a robust upper bound imposed by the perturbativity requirement for the considered mass configuration.

Subsequently, we investigate whether non-resonant annihilation alone is sufficient to reproduce the measured DM relic abundance when the gauge coupling is fixed at its perturbativity upper limit $g_V^{\rm max}=0.974$. The minimal gauge coupling values required to meet the relic density criterion $\Omega h^2= 0.12$ across the sampled range of DM masses are compiled in Table~\ref{tab:tchannel_gVmin}. Our analysis reveals that the minimal gauge coupling needed to satisfy the relic density constraint exhibits a monotonic increase with rising $m_V$. This behavior can be understood from the characteristic scaling relation governing the t-channel non-resonant annihilation cross section, $\langle\sigma v\rangle_{t}\propto g_V^{4}/m_V^{2}$. For elevated values of $m_V$, a proportionally higher $g_V$ is required to offset the associated phase-space suppression. At $m_V=700$~GeV, the derived minimal coupling $g_V^{\rm min}=0.817$ remains comfortably below the perturbativity bound. The critical crossover occurs at $m_V\simeq 1003$~GeV, beyond which $g_V^{\rm min}$ surpasses the admissible perturbativity limit.

\begin{table}[t]
	\centering
	\caption{Minimal values of the gauge coupling $g_V$ required to satisfy the DM relic density constraint for the pure non-resonant annihilation scenario, evaluated under the benchmark configuration with $m_{h_2}=100$~GeV and $\sin\alpha=0.01$. The perturbativity upper bound $g_V^{\rm max}=0.974$ eliminates nearly all viable parameter solutions corresponding to $m_V\gtrsim 1$~TeV.}
	\label{tab:tchannel_gVmin}
	\begin{tabular}{c|c|c}
		\hline
		$M_V$ [GeV] & $g_V^{\rm min}( \Omega h^2=0.12)$ & Viable? ($g_V^{\rm min}<0.974$) \\
		\hline
		200  & 0.475 & \checkmark \\
		500  & 0.696 & \checkmark \\
		700  & 0.817 & \checkmark \\
		1000 & 0.972 & \checkmark \\
		2000 & 1.368 & $\times$ \\
		4000 & 1.934 & $\times$ \\
		\hline
	\end{tabular}
\end{table}

This result exhibits a marked contrast with the s-channel resonant annihilation mechanism analyzed in Sec.~5.1. In that resonant scenario, the Breit–Wigner enhancement enables the observed relic abundance to be reproduced with a gauge coupling of order $g_V\sim\mathcal{O}(10^{-2})$, even for values of $m_{h_2}$ extending up to the multi-TeV scale. The non-resonant annihilation regime relies entirely on the magnitude of $g_V$ to compensate for the $1/m_V^{2}$ suppression of the annihilation cross section. The perturbativity constraint consequently confines the viable parameter space of the non-resonant mechanism to $m_V\lesssim 1$~TeV; beyond this threshold, the gauge coupling required to satisfy the relic density criterion can no longer be maintained within the range consistent with perturbative validity. We therefore conclude that, for the benchmark configuration $m_{h_2}=100$~GeV and $\sin\alpha=0.01$, the non-resonant parameter space is subject to stringent theoretical constraints, remaining only marginally viable at the onset of the heavy DM mass regime, prior to being fully excluded by the combined action of relic density requirements and perturbativity bounds.

\section{Conclusions}
The present study reports a systematic dedicated analysis of the heavy-mass parameter space for VDM realizations embedded within the $U(1)_X$ gauged scalar Higgs portal model. This model is parameterized by four independent free parameters: the VDM mass $m_V$, the dark scalar sector mass $m_{h_2}$, the scalar sector mixing angle $\sin\alpha$, and the dark gauge coupling $g_V$. We enforce a combined set of experimental and theoretical constraints, comprising the Planck upper limit on the DM relic abundance, the 2025 LZ spin-independent direct detection exclusion bound, the 14-year Fermi-LAT dSph $\gamma$-ray observational constraint, as well as a two-loop perturbativity criterion mandating that all scalar quartic couplings maintain values below $2\pi$ over the full energy range extending up to the Planck scale, to delineate the viable parameter space of the model within the multi-TeV region.

Subject to the complete set of applied experimental constraints, two distinct annihilation mechanisms are capable of accounting for the cosmologically observed DM relic abundance. Within the $s$-channel resonant parameter region defined by the condition $m_V \approx m_{h_2}/2$, the Breit–Wigner resonant enhancement enables the observed relic density to be reproduced with a gauge coupling as small as $g_V \sim \mathcal{O}(10^{-2})$. Upon incorporating the perturbativity constraint, we demonstrate that this resonant mechanism remains fully viable up to $m_{h_2} \simeq 2.43$ TeV for a fixed scalar mixing angle $\sin\alpha = 0.05$, and its viable parameter reach can be extended to $m_{h_2} \simeq 12.15$ TeV by reducing the mixing angle to $\sin\alpha = 0.01$. The scalar mixing angle is subject to an upper bound imposed by the perturbative validity requirement for the Higgs quartic coupling $\lambda_H$: for any specified value of $\sin\alpha$, there exists a critical maximum $m_{h_2}^{\max}$ beyond which the Higgs quartic coupling is driven past the perturbative validity threshold, irrespective of the configuration of other free model parameters.

In the non-resonant parameter regime defined by $m_{h_2}\lesssim m_V$, the t/u-channel-mediated annihilation of dark vector pairs into dark scalar pairs represents the dominant process governing the thermally averaged annihilation cross section. This channel can reproduce the DM relic abundance set by Planck measurements at elevated gauge coupling values, with the required coupling magnitude exhibiting no dependence on the scalar mixing angle. Nevertheless, the gauge coupling required to match the observed relic abundance exceeds the perturbativity bound $g_V^{\rm max}\simeq 0.974$ for all DM masses exceeding approximately $1$~TeV. Accordingly, this annihilation mechanism is almost entirely excluded across the heavy-mass regime, with only a narrow marginally allowed parameter window persisting at the low-mass boundary of this regime, which is ultimately ruled out under the combined constraints of relic density consistency and perturbativity across the full scope of the investigated parameter space.

To summarize, the phenomenologically viable parameter space of the VDM model within the multi-TeV mass regime is jointly bounded by complementary theoretical consistency constraints and experimental observational bounds. The perturbativity requirement for the scalar sector, enforced via two-loop RGE calculations, constitutes the decisive constraint that truncates the allowed parameter extents of both the $s$-channel resonant corridor and the $t/u$-channel dominated non-resonant region, thereby establishing a robust upper bound on the mass scales permittable within this DM framework. Notwithstanding the stringent perturbativity constraints applied across our analysis, we find that the model retains a sizable viable parameter space spanning the broad DM mass interval $1\,\text{TeV} \leq m_V \leq 10\,\text{TeV}$, provided the scalar mixing angle $\alpha$ is sufficiently small. This surviving viable parameter space is governed by the resonant annihilation mechanism, with the associated gauge coupling $g_V$ taking values in the range from $10^{-1}$ to $10^{-2}$.



\begin{thebibliography}{99}

\bibitem{Hambye:2009}     
  T.~Hambye,
  ``Hidden vector dark matter,''
  \href{https://doi.org/10.1088/1126-6708/2009/01/028}{J.\ High Energy Phys.\ {\bf 01}, 028 (2009) [arXiv:0811.0172 [hep-ph]]}.

\bibitem{Lebedev:2012}    
  O.~Lebedev, H.~M.~Lee and Y.~Mambrini,
  ``Vector Higgs-portal dark matter and the invisible Higgs,''
  \href{https://doi.org/10.1016/j.physletb.2012.01.029}{Phys.\ Lett.\ B {\bf 707}, 570--576 (2012) [arXiv:1111.4482 [hep-ph]]}.

\bibitem{Farzan:2012}     
  Y.~Farzan and A.~R.~Akbarieh,
  ``VDM: a model for vector dark matter,''
  \href{https://doi.org/10.1088/1475-7516/2012/10/026}{J.\ Cosmol.\ Astropart.\ Phys.\ {\bf 10}, 026 (2012) [arXiv:1207.4272 [hep-ph]]}.

\bibitem{Ko:2014}         
  P.~Ko, W.-I.~Park and Y.~Tang,
  ``Higgs portal vector dark matter for GeV scale $\gamma$-ray excess from
  Galactic Center,''
  \href{https://doi.org/10.1088/1475-7516/2014/09/013}{J.\ Cosmol.\ Astropart.\ Phys.\ {\bf 09}, 013 (2014) [arXiv:1404.5257 [hep-ph]]}.
  
\bibitem{Baek:2013}       
  S.~Baek, P.~Ko, W.-I.~Park and E.~Senaha,
  ``Higgs portal vector dark matter: revisited,''
  \href{https://doi.org/10.1007/JHEP05(2013)036}{J.\ High Energy Phys.\ {\bf 05}, 036 (2013) [arXiv:1212.2131 [hep-ph]]}.

\bibitem{Duch:2015}       
  M.~Duch, B.~Grzadkowski and M.~McGarrie,
  ``A stable Higgs portal with vector dark matter,''
  \href{https://doi.org/10.1007/JHEP09(2015)162}{J.\ High Energy Phys.\ {\bf 09}, 162 (2015) [arXiv:1506.08805 [hep-ph]]}.

  
    \bibitem{Ayazi:2023}      
  S.~Y.~Ayazi and M.~Hosseini,
  ``W-boson mass anomaly and vacuum structure in vector dark matter model with a
  singlet scalar mediator,''
  \href{https://doi.org/10.1142/S0217751X2340002X}{Int.\ J.\ Mod.\ Phys.\ A {\bf 38}, 2340002 (2023) [arXiv:2206.11041 [hep-ph]]}.
  
  \bibitem{Chao:2015}       
  W.~Chao,
  ``First order electroweak phase transition triggered by the Higgs portal vector
  dark matter,''
  \href{https://doi.org/10.1103/PhysRevD.92.015025}{Phys.\ Rev.\ D {\bf 92}, 015025 (2015) [arXiv:1412.3823 [hep-ph]]}.
  
  \bibitem{Hashino:2018}    
  K.~Hashino, M.~Kakizaki, S.~Kanemura, P.~Ko and T.~Matsui,
  ``Gravitational waves from first order electroweak phase transition in models
  with the $U(1)_X$ gauge symmetry,''
  \href{https://doi.org/10.1007/JHEP06(2018)088}{J.\ High Energy Phys.\ {\bf 06}, 088 (2018) [arXiv:1802.02947 [hep-ph]]}.
  
  \bibitem{Ayazi:2019}      
  S.~Y.~Ayazi and A.~Mohamadnejad,
  ``Conformal vector dark matter and strongly first-order electroweak phase
  transition,''
  \href{https://doi.org/10.1007/JHEP03(2019)181}{J.\ High Energy Phys.\ {\bf 03}, 181 (2019) [arXiv:1901.04168 [hep-ph]]}.
  
\bibitem{Arcadi:2020}     
  G.~Arcadi, A.~Djouadi and M.~Kado,
  ``The Higgs-portal for vector dark matter and the effective field theory
  approach: a reappraisal,''
  \href{https://doi.org/10.1016/j.physletb.2020.135427}{Phys.\ Lett.\ B {\bf 805}, 135427 (2020) [arXiv:2001.10750 [hep-ph]]}.

\bibitem{Arcadi:2024}     
  G.~Arcadi, J.~C.~Criado and A.~Djouadi,
  ``Iteration on the Higgs portal for vector dark matter and its effective field
  theory description,''
  \href{https://doi.org/10.1103/PhysRevD.109.075005}{Phys.\ Rev.\ D {\bf 109}, 075005 (2024) [arXiv:2312.14052 [hep-ph]]}.

\bibitem{Falkowski:2015}  
  A.~Falkowski, C.~Gross and O.~Lebedev,
  ``A second Higgs from the Higgs portal,''
  \href{https://doi.org/10.1007/JHEP05(2015)057}{J.\ High Energy Phys.\ {\bf 05}, 057 (2015) [arXiv:1502.01361 [hep-ph]]}.

\bibitem{Djouadi:2012}    
  A.~Djouadi, O.~Lebedev, Y.~Mambrini and J.~Quevillon,
  ``Implications of LHC searches for Higgs-portal dark matter,''
  \href{https://doi.org/10.1016/j.physletb.2012.01.062}{Phys.\ Lett.\ B {\bf 709}, 65--69 (2012) [arXiv:1112.3299 [hep-ph]]}.

\bibitem{Baek:2014}       
  S.~Baek, P.~Ko and W.-I.~Park,
  ``Invisible Higgs decay width vs.\ dark matter direct detection cross
  section in Higgs portal dark matter models,''
  \href{https://doi.org/10.1103/PhysRevD.90.055014}{Phys.\ Rev.\ D {\bf 90}, 055014 (2014) [arXiv:1405.3530 [hep-ph]]}.
  
\bibitem{Glaus:2019}      
S.~Glaus, M.~M\"{u}hlleitner, J.~M\"{u}ller, S.~Patel and R.~Santos,
``Electroweak corrections to dark matter direct detection in a vector dark
matter model,''
\href{https://doi.org/10.1007/JHEP10(2019)152}{J.\ High Energy Phys.\ {\bf 10}, 152 (2019) [arXiv:1908.09249 [hep-ph]]}.

\bibitem{Amiri:2023}      
A.~Amiri, B.~D\'{i}az S\'{a}ez and K.~Ghorbani,
``(sub)GeV dark matter in the $U(1)_X$ Higgs portal model,''
\href{https://doi.org/10.1016/j.physletb.2023.138119}{Phys.\ Lett.\ B {\bf 844}, 138119 (2023) [arXiv:2209.11723 [hep-ph]]}.

\bibitem{Ayazi:2024}      
S.~Y.~Ayazi, M.~Hosseini, S.~Paktinat Mehdiabadi and R.~Rouzbehi,
``Vector dark matter and LHC constraints, including a 95 GeV light Higgs
boson,''
\href{https://doi.org/10.1103/PhysRevD.110.055004}{Phys.\ Rev.\ D {\bf 110}, 055004 (2024) [arXiv:2405.01132 [hep-ph]]}.

\bibitem{Beniwal:2016}    
A.~Beniwal, F.~Rajec, C.~Savage, P.~Scott, C.~Weniger, M.~White and
A.~G.~Williams,
``Combined analysis of effective Higgs portal dark matter models,''
\href{https://doi.org/10.1103/PhysRevD.93.115016}{Phys.\ Rev.\ D {\bf 93}, 115016 (2016) [arXiv:1512.06458 [hep-ph]]}.

\bibitem{Shaikh:2026}     
  H.~Shaikh and M.~Di~Mauro,
  ``Vector Higgs-portal dark matter: How UV completion reopens viable
  parameter space,''
  \href{https://journals.aps.org/prd/abstract/10.1103/k7dg-mg3p}{Phys.\ Rev.\ D {\bf 113}, 115049 (2026)}
  [arXiv:2603.11233 [hep-ph]].

\bibitem{Baek:2022}       
  S.~Baek, P.~Ko and W.-I.~Park,
  ``Addendum to ``Invisible Higgs decay width versus dark matter direct
  detection cross section in Higgs portal dark matter models'',''
  \href{https://doi.org/10.1103/PhysRevD.105.015007}{Phys.\ Rev.\ D {\bf 105}, 015007 (2022) [arXiv:2112.11983 [hep-ph]]}.

\bibitem{Degrassi:2012}   
  G.~Degrassi, S.~Di~Vita, J.~Elias-Mir\'o, J.~R.~Espinosa,
  G.~F.~Giudice, G.~Isidori and A.~Strumia,
  ``Higgs mass and vacuum stability in the Standard Model at NNLO,''
  \href{https://doi.org/10.1007/JHEP08(2012)098}{J.\ High Energy Phys.\ {\bf 08}, 098 (2012) [arXiv:1205.6497 [hep-ph]]}.

\bibitem{Buttazzo:2013}   
  D.~Buttazzo, G.~Degrassi, P.~P.~Giardino, G.~F.~Giudice, F.~Sala,
  A.~Salvio and A.~Strumia,
  ``Investigating the near-criticality of the Higgs boson,''
  \href{https://doi.org/10.1007/JHEP12(2013)089}{J.\ High Energy Phys.\ {\bf 12}, 089 (2013) [arXiv:1307.3536 [hep-ph]]}.

\bibitem{Planck:2018}     
{\sc Planck Collaboration}, N.~Aghanim {\it et al.},
``Planck 2018 results. VI. Cosmological parameters,''
\href{https://doi.org/10.1051/0004-6361/201833910}{Astron.\ Astrophys.\ {\bf 641}, A6 (2020) [Erratum: Astron.\ Astrophys.\ {\bf 652}, C4 (2021)] [arXiv:1807.06209 [astro-ph.CO]]}.

\bibitem{PandaX:2025}     
{\textsc PandaX Collaboration}, Z.~Bo {\it et al.},
``Dark Matter Search Results from 1.54 Tonne$\cdot$Year Exposure of
PandaX-4T,''
\href{https://journals.aps.org/prl/abstract/10.1103/PhysRevLett.134.011805}{Phys.\ Rev.\ Lett.\ {\bf 134}, 011805 (2025)}
[\href{https://arxiv.org/abs/2408.00664}{arXiv:2408.00664 [hep-ex]}].

\bibitem{XENONnT:2025}    
{\textsc XENON Collaboration}, E.~Aprile {\it et al.},
``WIMP Dark Matter Search Using a 3.1 Tonne-Year Exposure of the XENONnT
Experiment,''
\href{https://journals.aps.org/prl/abstract/10.1103/msw4-t342}{Phys.\ Rev.\ Lett.\ {\bf 135}, 221003 (2025)}
[\href{https://arxiv.org/abs/2502.18005}{arXiv:2502.18005 [hep-ex]}].

\bibitem{LZ:2025}         
{\textsc LZ Collaboration}, J.~Aalbers {\it et al.},
``Dark Matter Search Results from 4.2 Tonne-Years of Exposure of the
LUX-ZEPLIN (LZ) Experiment,''
\href{https://journals.aps.org/prl/abstract/10.1103/4dyc-z8zf}{Phys.\ Rev.\ Lett.\ {\bf 135}, 011802 (2025)}
[\href{https://arxiv.org/abs/2410.17036}{arXiv:2410.17036 [hep-ex]}].

\bibitem{FermiLAT:2024}   
  A.~McDaniel, M.~Ajello, C.~M.~Karwin, M.~Di~Mauro,
  A.~Drlica-Wagner and M.~A.~S\'{a}nchez-Conde,
  ``Legacy Analysis of Dark Matter Annihilation from the Milky Way Dwarf
  Spheroidal Galaxies with 14 Years of Fermi-LAT Data,''
  \href{https://arxiv.org/abs/2311.04982}{Phys.\ Rev.\ D {\bf 109}, 063024 (2024) [arXiv:2311.04982 [astro-ph.HE]]}.

 \bibitem{Fermi-LAT:2015kyq}         
{\textsc Fermi-LAT Collaboration}, M.~Ackermann {\it et al.},
``Updated Search for Spectral Lines from Galactic Dark Matter Interactions with Pass 8 Data from the Fermi Large Area Telescope,''
\href{https://journals.aps.org/prd/abstract/10.1103/PhysRevD.91.122002}{Phys.\ Rev.\ D\ {\bf 91}, 122002 (2015)}
[\href{https://arxiv.org/abs/1506.00013}{arXiv:1506.00013 [astro-ph.HE]}].

 \bibitem{DAMPE2022}         
{\textsc DAMPE Collaboration}, F.~Alemanno {\it et al.},
``Search for Gamma-Ray Spectral Lines with the DArk Matter Particle Explorer,''
\href{https://www.sciencedirect.com/science/article/pii/S2095927321007672}{Science\ Bulletin\ {\bf 67}, 679 (2022)}
[\href{https://arxiv.org/abs/2112.08860}{arXiv:2112.08860 [astro-ph.HE]}].

 \bibitem{ATLAS:2023tkt}         
{\textsc ATLAS Collaboration}, G.~Aad {\it et al.},
``Combination of searches for invisible decays of the Higgs boson using $139\,\text{fb}^{-1}$ of proton-proton collision data at $\sqrt{s}=13$~TeV collected with the ATLAS experiment,''
\href{https://www.sciencedirect.com/science/article/pii/S0370269323002976}{Phys.\ Lett.\ B\ {\bf 842}, 137963 (2023)}
[\href{http://arxiv.org/abs/2301.10731} {arXiv:2301.10731 [hep-ex]}].

\bibitem{Frandsen:2023}
  M.~T.~Frandsen, M.~Heikinheimo, M.~E.~Thing, K.~Tuominen,
  M.~Rosenlyst,
  ``Vector dark matter in supercooled Higgs portal models,''
  \href{https://doi.org/10.1103/PhysRevD.108.015033}{Phys.\ Rev.\ D {\bf 108}, 015033 (2023) [arXiv:2301.00041 [hep-ph]]}.
  
  \bibitem{Kolb:1990}       
  E.~W.~Kolb and M.~S.~Turner,
  \textit{The Early Universe} (CRC Press, 1990).

\bibitem{MicrOMEGAs:2023} 
  G.~Alguero, G.~B\'{e}langer, F.~Boudjema, S.~Chakraborti,
  A.~Goudelis, S.~Kraml, A.~Mjallal and A.~Pukhov,
  ``micrOMEGAs 6.0: $N$-component dark matter,''
  \href{https://doi.org/10.1016/j.cpc.2024.109133}{Comput.\ Phys.\ Commun.\ {\bf 299}, 109133 (2024) [arXiv:2312.14894 [hep-ph]]}.

\bibitem{SARAH4}          
  F.~Staub,
  ``SARAH 4: A tool for (not only) SUSY model builders,''
  \href{https://arxiv.org/abs/1309.7223}{Comput.\ Phys.\ Commun.\ {\bf 185}, 1773--1790 (2014) [arXiv:1309.7223 [hep-ph]]}.

\bibitem{SARAH3}          
  F.~Staub,
  ``SARAH 3.2: Dirac Gauginos, UFO output, and more,''
  \href{https://arxiv.org/abs/1207.0906}{Comput.\ Phys.\ Commun.\ {\bf 184}, 1792--1809 (2013) [arXiv:1207.0906 [hep-ph]]}.

\bibitem{SARAH:Unitarity} 
  M.~D.~Goodsell and F.~Staub,
  ``Unitarity constraints on general scalar couplings with SARAH,''
  \href{https://arxiv.org/abs/1805.07306}{Eur.\ Phys.\ J.\ C {\bf 78}, 649 (2018) [arXiv:1805.07306 [hep-ph]]}.

\bibitem{SARAH:Goldstone} 
  J.~Braathen, M.~D.~Goodsell and F.~Staub,
  ``Supersymmetric and non-supersymmetric models without catastrophic Goldstone
  bosons,''
  \href{https://arxiv.org/abs/1706.05372}{Eur.\ Phys.\ J.\ C {\bf 77}, 757 (2017) [arXiv:1706.05372 [hep-ph]]}.

\bibitem{SPheno:2003}     
  W.~Porod,
  ``SPheno, a program for calculating supersymmetric spectra, SUSY particle
  decays and SUSY particle production at $e^+e^-$ colliders,''
  \href{https://arxiv.org/abs/hep-ph/0301101}{Comput.\ Phys.\ Commun.\ {\bf 153}, 275--315 (2003) [arXiv:hep-ph/0301101]}.

\bibitem{SPheno:2012}     
  W.~Porod and F.~Staub,
  ``SPheno 3.1: Extensions including flavour, CP-phases and models beyond
  the MSSM,''
  \href{https://arxiv.org/abs/1104.1573}{Comput.\ Phys.\ Commun.\ {\bf 183}, 2458--2469 (2012) [arXiv:1104.1573 [hep-ph]]}.
  
  \bibitem{Schumann:2019}   
  M.~Schumann,
  ``Direct detection of WIMP dark matter: concepts and status,''
  \href{https://arxiv.org/abs/1903.03026}{J.\ Phys.\ G {\bf 46}, 103003 (2019) [arXiv:1903.03026 [astro-ph.CO]]}.
  
\end{thebibliography}
\end{document}